Comments on "Climbing Escher's stairs: A way to approximate stability landscapes in multidimensional systems"

Jingmeng Cui[a*], Anna Lichtwarck-Aschoff[a], and Fred Hasselman[b]

[a]*Faculty of Behavioural and Social Sciences, University of Groningen, Groningen, the Netherlands;* [b]*Behavioural Science Institute, Radboud University, Nijmegen, the Netherlands.*

*Corresponding author

Jingmeng Cui, jingmeng.cui@rug.nl, Grote Kruisstraat 2/1, 9712 TS Groningen, the Netherlands

ORCIDs:

Jingmeng Cui: 0000-0003-3421-8457

Anna Lichtwarck-Aschoff: 0000-0002-4365-1538

Fred Hasselman: 0000-0003-1384-8361

A paper published in *PLOS Computational Biology* in 2020 [1] proposed a new method to construct a potential landscape function for multivariate systems, along with an R package *rolldown* (now renamed as *waydown* [2]). The approach proposed by the authors is based on the local decomposition of the Jacobian matrix. As shown by the authors, the potential function *V* for a multivariate dynamic system only exists if the Jacobian matrix is symmetric. By taking the symmetric part of the Jacobian in the path integral, the authors claim that the algorithm they proposed was "an approximation of the above-mentioned Helmholtz decomposition, i.e., to decompose differential equations as the sum of a gradient and a non-gradient, divergence-free part", and specifically, in the supplementary materials, the authors claim that "because we are building our potential neglecting the non-gradient part of our vector field, we know that our results will converge to the same solution regardless of the chosen path." We sincerely appreciate the efforts made by the authors in pursuit of a computationally efficient method for calculating the potential landscape functions. However, in this Formal Comment, we would like to point out an important limitation of this method that we believe warrants discussion, which is that even the potential landscape function constructed from the *gradient* part defined in this article is path-dependent. Moreover, in the implementation provided by the authors, the decomposition method was ineffective, and removing the decomposition does not alter the output of the algorithm.

To illustrate those points, we first consider the existence condition of the potential function. As shown by the authors, the strictly defined potential function for a two-dimensional dynamic system only exists when the crossed derivatives of the dynamic functions are equal. If the dynamic function can be decomposed into two parts, in which one of them satisfies this condition, then the potential function of this gradient part can be calculated and serves as a generalized potential function for the original dynamic system. This approach was applied successfully in previous studies (e.g., [3,4]), as mentioned by the authors of [1]. However, in the decomposition method proposed in [1], the decomposition was only applied to the Jacobian matrix, but not to the original dynamic functions. As a result, no gradient function is yielded from the decomposition, which makes the output of the algorithm still path-dependent. We illustrate this point in a concrete example. Consider the following dynamic system:

$$\begin{cases} \dfrac{dx}{dt} = f(x,y) = 0, \\ \dfrac{dy}{dt} = g(x,y) = x. \end{cases} \tag{1}$$

The Jacobian matrix of the system is given by the following:

$$J = \begin{bmatrix} \dfrac{\partial f}{\partial x} & \dfrac{\partial f}{\partial y} \\ \dfrac{\partial g}{\partial x} & \dfrac{\partial g}{\partial y} \end{bmatrix} = \begin{bmatrix} 0 & 0 \\ 1 & 0 \end{bmatrix}. \tag{2}$$

Using the method described in Eq (9) of [1], $J$ can be decomposed into the sum of the $J_{symm}$ and $J_{skew}$:

$$\begin{cases} J_{symm} = \begin{bmatrix} 0 & 0.5 \\ 0.5 & 0 \end{bmatrix}, \\ J_{skew} = \begin{bmatrix} 0 & -0.5 \\ 0.5 & 0 \end{bmatrix}. \end{cases} \tag{3}$$

To show the path-dependency of the integral, we consider two pathways to calculate $\Delta V$ from (0,0) to (1,1) using the method described in Eq (16-18) of [1]. To avoid the calculation error associated with grid size, here we consider the case with an infinitely dense grid. The first pathway is first along the *x*-axis, then along the *y*-axis:

$$\Delta V(1,1; 0,0) = \lim_{n \to \infty} \sum_{i=1}^{n} \Delta V\left(\frac{i}{n}, 0; \frac{i-1}{n}, 0\right) + \sum_{j=1}^{n} \Delta V\left(1, \frac{j}{n}; 1, \frac{j-1}{n}\right)$$

$$= \lim_{n \to \infty} \sum_{i=1}^{n} \left\{-\left[0, \frac{i-1}{n}\right]\left[\frac{1}{n}, 0\right]^T - \frac{1}{2}\left[\frac{1}{n}, 0\right]^T \begin{bmatrix} 0 & 0.5 \\ 0.5 & 0 \end{bmatrix}\left[\frac{1}{n}, 0\right]\right\}$$

$$+ \sum_{j=1}^{n} \left\{-[0,1]\left[0, \frac{1}{n}\right]^T - \frac{1}{2}\left[0, \frac{1}{n}\right]^T \begin{bmatrix} 0 & 0.5 \\ 0.5 & 0 \end{bmatrix}\left[0, \frac{1}{n}\right]\right\}$$

$$= \lim_{n \to \infty} \sum_{i=1}^{n} \{-0 - 0\} + \sum_{j=1}^{n} \left\{-\frac{1}{n} - 0\right\} = -1. \tag{4}$$

The second pathway is first along the *y*-axis, then along the *x*-axis:

$$\Delta V(1,1; 0,0) = \lim_{n \to \infty} \sum_{i=1}^{n} \Delta V\left(0, \frac{i}{n}; 0, \frac{i-1}{n}\right) + \sum_{j=1}^{n} \Delta V\left(\frac{j}{n}, 1; \frac{j-1}{n}, 1\right)$$

$$= \lim_{n \to \infty} \sum_{i=1}^{n} \left\{-[0,0]\left[0, \frac{1}{n}\right]^T - \frac{1}{2}\left[0, \frac{1}{n}\right]^T \begin{bmatrix} 0 & 0.5 \\ 0.5 & 0 \end{bmatrix}\left[0, \frac{1}{n}\right]\right\}$$

$$+ \sum_{j=1}^{n} \left\{-\left[0, \frac{j-1}{n}\right]\left[\frac{1}{n}, 0\right]^T - \frac{1}{2}\left[\frac{1}{n}, 0\right]^T \begin{bmatrix} 0 & 0.5 \\ 0.5 & 0 \end{bmatrix}\left[\frac{1}{n}, 0\right]\right\}$$

$$= \lim_{n \to \infty} \sum_{i=1}^{n} \{-0 - 0\} + \sum_{j=1}^{n} \{-0 - 0\} = 0. \tag{5}$$

Comparing the results, we can see that even if we use the symmetric part of the Jacobian for calculation, the potential function is still path-dependent. This is because in each calculation step described in Eq (16) of [1], not only the Jacobian matrix but also the value of the *original dynamic functions*, $\vec{f}(\vec{x_0})$, was used. Therefore, the value of the original functions, which was not decomposed, also affects the final calculation.

Moreover, as described in the supplementary materials of [1], the authors mentioned that the integral calculation was always either along the *x*-axis or the *y*-axis in their algorithm. This renders the decomposition step ineffective because $J_{symm}$ and $J$ only differ in the antidiagonal elements, but the value of the quadratic form used in Eq (16) of [1], $-\frac{1}{2}\Delta\vec{x}^T J_{symm}(\vec{x_0})\Delta\vec{x}$, only depends on the diagonal elements when $\Delta\vec{x}$ is along either the *x*-axis or the *y*-axis. Therefore, the output of the algorithm is always the same as the mean of the path integrals along the two paths (i.e., first along the *x*-axis, then along the *y*-axis, and vice versa). In the supplementary materials of this Formal Comment, we show that after removing the decomposition step in the *waydown* package [1,2], the output remains identical.

Therefore, based on the theoretical analysis and examples we showed above, we conclude that the method proposed in [1] does not yield a potential function that is path-independent up to the level of the computational approximations. Specifically, if the paths are chosen either along the *x*-axis or the *y*-axis as implemented by the author of [1], the decomposition step proposed as the core of the algorithm does not have an influence on the output. This limitation may require the attention of potential readers and users, and researchers seeking a potential landscape calculation method that is not path-dependent may need to consider alternative algorithms. In [1], the authors provide a literature review of the existing methods of landscape construction, which can serve as general guidance. For those searching for implementations in R, there are several packages available to our knowledge. The *QPot* package [5,6], for example, provides a method based on path integrals and the ordered upwind method; the *simlandr* package [7–9] provides a method based on numeric estimation of the steady-state distribution and Wang's definition of the generalized potential landscape [10]; the *fitlandr* package [11,12] provides an implementation of the Bhattacharya method [13]. It should be noted, however, that those methods are more computationally intensive than the method in [1] and require substantially longer time of calculation.

# `waydown` Package without the Decomposition Step

Jingmeng Cui, Anna Lichtwarck-Aschoff, Fred Hasselman

In this file, we illustrate after removing the decomposition step in the `waydown` package, the output remains identical.

We use the test function from the vignette of the `waydown` package:

```r
bx <- 0.2
ax <- 0.125
kx <- 0.0625
rx <- 1

by <- 0.05
ay <- 0.1094
ky <- 0.0625
ry <- 1

n <- 4

# Dynamics
f <- function(x) {c(bx - rx*x[1] + ax/(kx + x[2]^n), by - ry*x[2] + ay/(ky + x[1]^n))}
```

## Landscape with decomposition

In this section, we use the original `waydown` package.

```r
xs <- seq(0, 4, by = 0.05)
ys <- seq(0, 4, by = 0.05)
result <- waydown::approxPot2D(f, xs, ys)
data <- expand.grid(X = xs, Y = ys)
data$V <- as.vector(result$V)
data$err <- as.vector(result$err)
# Input equilibrium points (calculated externally)
eqPoints <- data.frame(x_eq = c(0.213416, 0.559865, 2.19971),
                       y_eq = c(1.74417, 0.730558, 0.0546602),
                       equilibrium = factor(c('stable', 'unstable', 'stable')))

nbins <- 25
library(ggplot2)
plotV <- ggplot() +
         geom_tile(data = data, aes(x = X, y = Y, fill = V)) +
         geom_contour(data = data, aes(x = X, y = Y, z = V), colour = 'white', alpha = 0.5, bins = nbi
         geom_point(data = eqPoints, aes(x = x_eq, y = y_eq, color = equilibrium)) +
         coord_fixed() +
         scale_fill_gradientn(colours = colorRamps::matlab.like(nbins)) +
```

```
            xlab("x") + ylab("y") + ggtitle("Approximate potential") +
            theme_bw()
plotV
```

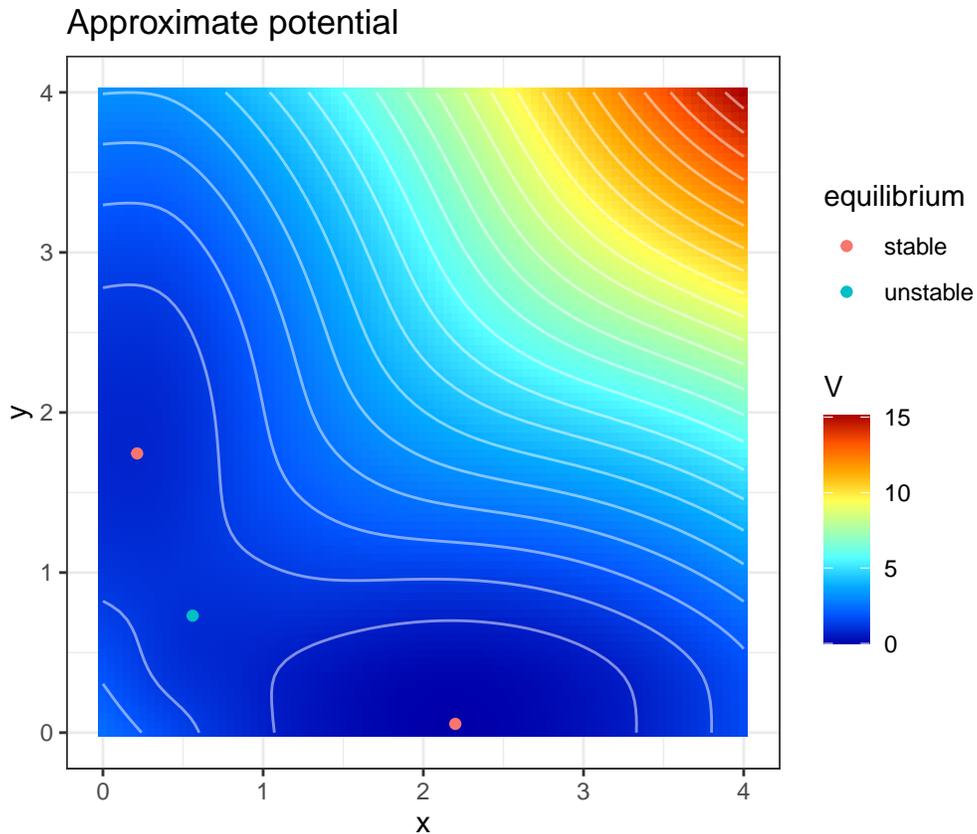

# Landscape withOUT decomposition

In this section, we rewrite the functions from the `waydown` package to remove the decomposition step.

```
deltaV2 <- function(f, x, x0, normType='f') {

  # Calculate the local Jacobian
  J0 <- numDeriv::jacobian(f, x0)

  # Perform the skew/symmetric decomposition
  J_symm <- Matrix::symmpart(J0)
  J_skew <- Matrix::skewpart(J0)

  # Use J_symm to estimate the difference in potential as 2nd order Taylor expansion
  #
  # Detailed information available at https://doi.org/10.1371/journal.pcbi.1007788
  dV <- as.numeric(
        -f(x0) %*% (x - x0) +   # Linear term
        -0.5 * t(x-x0) %*% J0 %*% (x - x0) # Quadratic term
```

```r
  )
  ##########################################
  ### Here I use J0 instead of J_symm  in waydown::deltaV2()
  ### This is the only place that we made a change.
  ##########################################

  # Use J_skew to estimate the relative error
  #
  # Detailed information available at https://doi.org/10.1371/journal.pcbi.1007788
  rel_err <- norm(J_skew, type = normType)/(norm(J_skew, type = normType) + norm(J_symm, type = normType
  
  # Return
  ls <- list(dV = dV, err = rel_err)
  return(ls)
}

approxPot2D2 <- function(f, xs, ys, V0 = 'auto', mode = 'mixed') {
  # Initialize
  V <- matrix(0, nrow = length(xs), ncol = length(ys))
  err <- matrix(0, nrow = length(xs), ncol = length(ys))
  
  # Assign initial value
  # The algorithm is a recursion relationship. It needs an initial potential at the first integration p
  if (V0 == 'auto') {
    V[1,1] <- 0 # Assign any, it will be overriden later
  } else {
    V[1,1] <- V0 # Assign the desired reference potential
  }
  
  # Compute
  # We first compute along the first column...
  for(i in 2:length(xs)) {
    temp <- deltaV2(f, c(xs[i], ys[1]), c(xs[i-1], ys[1]))
    V[i,1] <- V[i-1,1] + temp$dV
    err[i,1] <- temp$err
  }
  
  # ... and then along the first row...
  for(j in 2:length(ys)) {
    temp <- deltaV2(f, c(xs[1], ys[j]), c(xs[1], ys[j-1]))
    V[1,j] <- V[1,j-1] + temp$dV
    err[1,j] <- temp$err
  }
  
  # ... and last but not least, we fill the inside gaps
  for(i in 2:length(xs)) {
    for(j in 2:length(ys)) {
      
      if(mode == 'horizontal') { # Sweep horizontally
        
        temp <- deltaV2(f, c(xs[i], ys[j]), c(xs[i-1], ys[j]))
        V[i,j] <- V[i-1,j] + temp$dV
        err[i,j] <- temp$err
```

```r
      } else if(mode == 'vertical') { # Sweep vertically
        
        temp <- deltaV2(f, c(xs[i], ys[j]), c(xs[i], ys[j-1]))
        V[i,j] <- V[i,j-1] + temp$dV
        err[i,j] <- temp$err
        
      } else if(mode == 'mixed') { # Sweep in both directions, then take the mean
        
        temp_hor <- deltaV2(f, c(xs[i], ys[j]), c(xs[i-1], ys[j]))
        V_hor <- V[i-1,j] + temp_hor$dV
        temp_ver <- deltaV2(f, c(xs[i], ys[j]), c(xs[i], ys[j-1]))
        V_ver <- V[i,j-1] + temp_ver$dV
        V[i,j] <- mean(c(V_hor, V_ver))
        err[i,j] <- mean(c(temp_hor$err, temp_ver$err))
        
      } else {
        
        stop('Error: supported modes are horizontal (default), vertical and mixed')
        
      }
    }
  }
  
  if(V0 == 'auto') {
    V <- V - min(c(V)) # Make V_min = 0
  }
  
  return(list(V = V, err = err))
}
```

```r
result2 <- approxPot2D2(f, xs, ys)
data2 <- expand.grid(X = xs, Y = ys)
data2$V <- as.vector(result2$V)
data2$err <- as.vector(result2$err)
plotV2 <- ggplot() +
          geom_tile(data = data2, aes(x = X, y = Y, fill = V)) +
          geom_contour(data = data2, aes(x = X, y = Y, z = V), colour = 'white', alpha = 0.5, bins = nb
          geom_point(data = eqPoints, aes(x = x_eq, y = y_eq, color = equilibrium)) +
          coord_fixed() +
          scale_fill_gradientn(colours = colorRamps::matlab.like(nbins)) +
          xlab("x") + ylab("y") + ggtitle("Approximate potential") +
          theme_bw()
plotV2
```

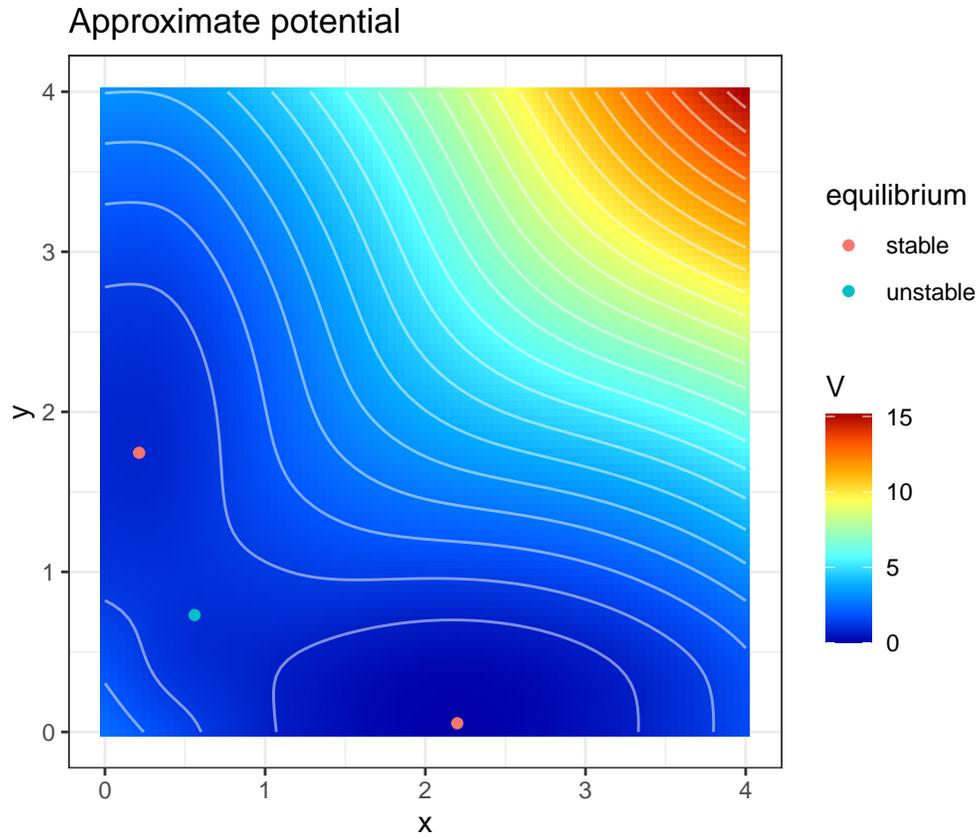

```
identical(result, result2)
```

```
## [1] TRUE
```

After removing the decomposition part in the algorithm, the result is exactly identical.

(The results are exactly identical, not approximately identical, because in each step $\Delta x$ is always along the axes, therefore only the diagonal elements of the Jacobian can have an effect. The diagonal elements are identical for the original Jacobian and the symmetric part.)

# Session information

```
sessionInfo()
```

```
## R version 4.3.2 (2023-10-31 ucrt)
## Platform: x86_64-w64-mingw32/x64 (64-bit)
## Running under: Windows 10 x64 (build 19045)
##
## Matrix products: default
##
##
## locale:
```

```
## [1] LC_COLLATE=English_United States.utf8
## [2] LC_CTYPE=English_United States.utf8
## [3] LC_MONETARY=English_United States.utf8
## [4] LC_NUMERIC=C
## [5] LC_TIME=English_United States.utf8
## 
## time zone: Europe/Berlin
## tzcode source: internal
## 
## attached base packages:
## [1] stats     graphics  grDevices utils     datasets  methods   base
## 
## other attached packages:
## [1] ggplot2_3.4.4
## 
## loaded via a namespace (and not attached):
##  [1] vctrs_0.6.4          cli_3.6.1            knitr_1.45
##  [4] rlang_1.1.2          xfun_0.41            highr_0.10
##  [7] generics_0.1.3       labeling_0.4.3       glue_1.6.2
## [10] isoband_0.2.7        colorspace_2.1-0     htmltools_0.5.7
## [13] fansi_1.0.5          scales_1.2.1         rmarkdown_2.25
## [16] grid_4.3.2           waydown_1.1.0        evaluate_0.23
## [19] munsell_0.5.0        tibble_3.2.1         fastmap_1.1.1
## [22] yaml_2.3.7           lifecycle_1.0.4      numDeriv_2016.8-1.1
## [25] compiler_4.3.2       dplyr_1.1.3          colorRamps_2.3.1
## [28] pkgconfig_2.0.3      rstudioapi_0.15.0    farver_2.1.1
## [31] lattice_0.21-9       digest_0.6.33        R6_2.5.1
## [34] tidyselect_1.2.0     utf8_1.2.4           pillar_1.9.0
## [37] magrittr_2.0.3       Matrix_1.6-1.1       withr_2.5.2
## [40] tools_4.3.2          gtable_0.3.4
```